# PubMed knowledge graph 2.0: Connecting papers, patents, and clinical trials in biomedical science

Jian Xu[1], Chao Yu[1], Jiawei Xu[2], Vetle I. Torvik[3], Jaewoo Kang[4], Mujeen Sung[5], Min Song[6], Yi Bu[7] & Ying Ding[2,8]

[1]School of Information Management, Sun Yat-sen University, Guangzhou, China.
[2]School of Information, University of Texas at Austin, Austin, TX, USA.
[3]School of Information Sciences, University of Illinois at Urbana-Champaign, Champaign, IL, USA.
[4]Department of Computer Science and Engineering, Korea University, Seoul, South Korea.
[5]School of Computing, Kyung Hee University, Seoul, South Korea.
[6]Department of Library and Information Science, Yonsei University, Seoul, South Korea.
[7]Department of Information Management, Peking University, Beijing, China
[8]Dell Medical School, University of Texas at Austin, Austin, TX, USA.

Corresponding authors: Jian Xu (issxj@mail.sysu.edu.cn), Ying Ding (ying.ding@austin.utexas.edu)

## Abstract

Papers, patents, and clinical trials are essential scientific resources in biomedicine, crucial for knowledge sharing and dissemination. However, these documents are often stored in disparate databases with varying management standards and data formats, making it challenging to form systematic and fine-grained connections among them. To address this issue, we construct PKG 2.0, a comprehensive knowledge graph dataset encompassing over 36 million papers, 1.3 million patents, and 0.48 million clinical trials in the biomedical field. PKG 2.0 integrates these dispersed resources through 482 million biomedical entity linkages, 19 million citation linkages, and 7 million project linkages. The construction of PKG 2.0 wove together fine-grained biomedical entity extraction, high-performance author name disambiguation, multi-source citation integration, and high-quality project data from the NIH Exporter. Data validation demonstrates that PKG 2.0 excels in key tasks such as author disambiguation and biomedical entity recognition. This dataset provides valuable resources for biomedical researchers, bibliometric scholars, and those engaged in literature mining.

## Background & Summary

The rapid expansion of digital data in scholarly domains has unlocked new horizons for exploring patterns characterizing the structure and evolution of science[1]. Academic papers, patents, and clinical trials are three distinct yet interconnected components in the realm of scholarly communication of medicine. Each serves a unique purpose and follows a specific structure, yet they all share the common goal of disseminating knowledge and findings.

Academic papers aim to further science through the dissemination of original research, often peer-reviewed to ensure validity and quality[2,3,4,5]. They provide detailed descriptions of research methods and results, serving as valuable references for other researchers. However, the impact of academic papers is often confined to the academic community, making it challenging to translate findings directly into practical applications. In contrast, patents offer significant advantages in protecting innovation as legal documents that establish intellectual property rights for new inventions and processes[6,7,8,9,10]. They can reveal technological trends and directions in industry research, providing insights into the commercialization of academic inquiries[11,12,13]. However, patents generally lack extensive experimental data, theoretical insights, and comprehensive discussions of scientific principles that are characteristic of papers and clinical trial studies, which can lead to a gap in the detailed understanding of technological developments[14,15,16,17,18]. Clinical trial records provide a comprehensive overview of all aspects of the clinical trials, producing study records including discussions and conclusions that supply context to the results. The results of clinical trials are crucial for regulatory approval and clinical adoption. These records, while rich in procedural detail and regulatory compliance, typically lack the exploratory scientific narratives and theoretical underpinnings that academic papers elaborate on[19,20,21]. Similarly, they do not delve into the technical specifications and potential commercial applications that are the focus of patents[22,23]. This absence of information can create a disconnect between the practical execution of clinical trials and the broader scientific and commercial context of medical research[24,25,26,27,28,29].

While each type of scholarly document—be it academic papers, patents, or clinical trial records—serves a unique function within the scientific landscape, the potential synergies between them are often not fully realized. Establishing a more integrated approach to these diverse knowledge repositories could significantly enrich our comprehension of the scientific ecosystem. By fostering a more interconnected framework, we can facilitate a multidimensional perspective that enhances innovation, accelerates the translation of research into practical applications, and promotes a more collaborative environment for scientific discovery[30,31,32,33]. This holistic view would not only bridge the existing gaps but also pave the way for a more cohesive and progressive scientific community. Some studies have obtained interesting findings using data from various types of literature. For example, a study on Bee Pollen has effectively utilized academic papers, patents, and clinical trials to present a comprehensive view of its health benefits[34]; Meyer et al.[35] used both patent and paper literature for scientometric analysis and found that inventor-authors tend to be more productive and academically influential than ordinary authors; Thelwall et al.[36] conducted citation analysis on clinical trials and papers and found that articles cited by clinical trials are usually more likely to have long-term impact.

The efforts in constructing knowledge graphs from academic literature can be categorized as follows:

(1) Fine-grained single-type literature knowledge graph. This category refers to works that utilize only one type of scholarly output, either academic papers, patents, or clinical trials. Each type provides unique insights and data that contribute to the construction of knowledge graphs. Chen et al.[37] constructed a publicly available Clinical Trial Knowledge Graph (CTKG), including nodes representing medical entities in clinical trials (such as studies, drugs, and conditions) and edges representing relationships between these entities (such as drugs used in studies). Patent data is often used to build engineering knowledge graphs[38,39]. Zou et al.[40] constructed a patent-based knowledge graph, patent-KG, to represent knowledge facts in engineering design patents. Knowledge graphs like the Clinical Trial Knowledge Graph (CTKG) and patent-KG serve as specialized repositories that capture the nuances of their respective domains. CTKG, for instance, provides a structured representation of clinical trials, offering insights into the relationships between medical entities such as drugs and conditions. This specificity is invaluable for researchers focusing on clinical outcomes and drug efficacy. Similarly, patent-KG encapsulates the inventive steps and technical details of engineering design patents, illuminating the pathways of technological innovation. However, because they are independent knowledge graphs summarized from the single-type literature, they cannot provide the association of knowledge among theoretical, clinical, and commercial research. Consequently, scientometrics analysis and knowledge discovery research based on such a graph are subject to certain constraints.

(2) Coarse-grained Integrated multi-type academic graph. This type of graph involves the fusion of different types of scholarly outputs. SciSciNet[41], a large-scale open data lake for scientific research, contains a database that captures scientific publications, researchers, and institutions and tracks their connections to related entities, including citations to patents, scientific publications in clinical trials, and social media mentions. The Dimensions database[42] indexes papers and their citations and uniquely connects publications to their related grants, funding agencies, patents, and clinical trials. Bibliometric networks like Dimensions and SciSciNet offer a more holistic view of scholarly communication by integrating diverse types of literature. However, while these knowledge graphs have significantly advanced our understanding of scholarly communication, they lack a more refined, multi-perspective linking approach between different types of scientific literature and evaluative entities.

Overall, the current knowledge graph, whether it is a fine-grained knowledge graph based on knowledge entities like gene, drug, and disease or a coarse-grained knowledge graph based on evaluation entities like paper, author, and journal, lacks multi-perspective and fine-grained comprehensive links to create stronger associations between different knowledge. This limitation prevents the full utilization of the interconnectedness of various types of scholarly outputs, thereby restricting the potential for more holistic scientometrics analysis and knowledge discovery.

In 2020, we developed a PubMed knowledge graph (referred to as PKG 1.0)[43] based on the PubMed 2019 dataset (https://pubmed.ncbi.nlm.nih.gov/). PKG 1.0 created connections among the bio-entities, authors, articles, affiliations, and funding. It facilitates measuring scholarly impact, knowledge usage, and knowledge transfer, as well as profiling authors and organizations based on their connections with bio-entities. Recognized for its utility, PKG 1.0 was supplemented by Scientific Data with "a detailed open access model of the PubMed literature", enhancing its reach and applicability[44]. Over the subsequent years, PKG 1.0 has been utilized across different research domains such as artificial intelligence[45,46], information retrieval[47,48], and scientometrics[49,50]. As shown in Table 1, PKG 1.0 has been cited more than 100 times in Web of Science, with the citation fields predominantly in Computer Science, Information Science Library Science, and Mathematical Computational Biology.

Table. 1 Distribution of Citation Fields of PKG 1.0 in Web of Science.

| Field | Number of citations | Example of specific studies (Paper title) | Purpose of using PKG 1.0 |
|---|---|---|---|
| Computer Science | 83 | SPACES: Explainable Multimodal AI for Active Surveillance, Diagnosis, and Management of Adverse Childhood Experiences (ACEs)[51] | Constructing an explainable multimodal AI |
| Information Science Library Science | 52 | Building the COVID-19 Portal by Integrating Literature, Clinical Trials, and Knowledge Graphs[52] | Building an information retrieval system |
| Mathematical Computational Biology | 38 | Analyzing knowledge entities about COVID-19 using entitymetrics[53] | Analyzing knowledge entities about COVID-19 |
| Mathematics | 27 | A graph neural network-based node classification model on class-imbalanced graph data[54] | Building a node classification model |
| Health Care Sciences Services | 23 | Digital Personal Health Coaching Platform for Promoting Human Papillomavirus Infection Vaccinations and Cancer Prevention: Knowledge Graph-Based Recommendation System[55] | Constructing a digital personal health coaching platform |

The table illustrates the distribution of citation fields of PKG 1.0 within the Web of Science database. Each line represents a different research field, with the number of citations, example of specific studies, and purpose of using PKG 1.0. Computer Science leads with 83 citations, followed by Information Science Library Science with 52 citations. Other notable fields include Mathematical Computational Biology, Mathematics, and Health Care Sciences Services. This data was queried from Web of Science on January 06, 2025[56].

However, PKG 1.0 only focused on paper data, which presents limitations when compared to integrating papers, patents, and clinical trials. This focus restricts the ability to support comprehensive research across different types of academic literature. The connections between these three types of literature are also crucial for a holistic understanding of knowledge transfer and innovation pathways. The necessity for PKG 2.0 arises from the ever-growing complexity of biomedical research and the increasing interdependence of scientific literature. PKG 2.0 is designed to address these challenges by integrating papers, patents, and clinical trials as core data, establishing fine-grained connections between these document types through bioentities, citations, and projects. Additionally, based on this linked dataset, we have conducted supplementary and extended works, including institution disambiguation, calculating commonly used measurements of authors, integrating citation data from multiple sources, and supplementing detailed information of journals annually (see Figure 1 for more details). By integrating these diverse sources of knowledge, PKG 2.0 aims to provide a holistic view of the research landscape, enabling researchers to uncover patterns and insights that were previously obscured by the siloed nature of data.

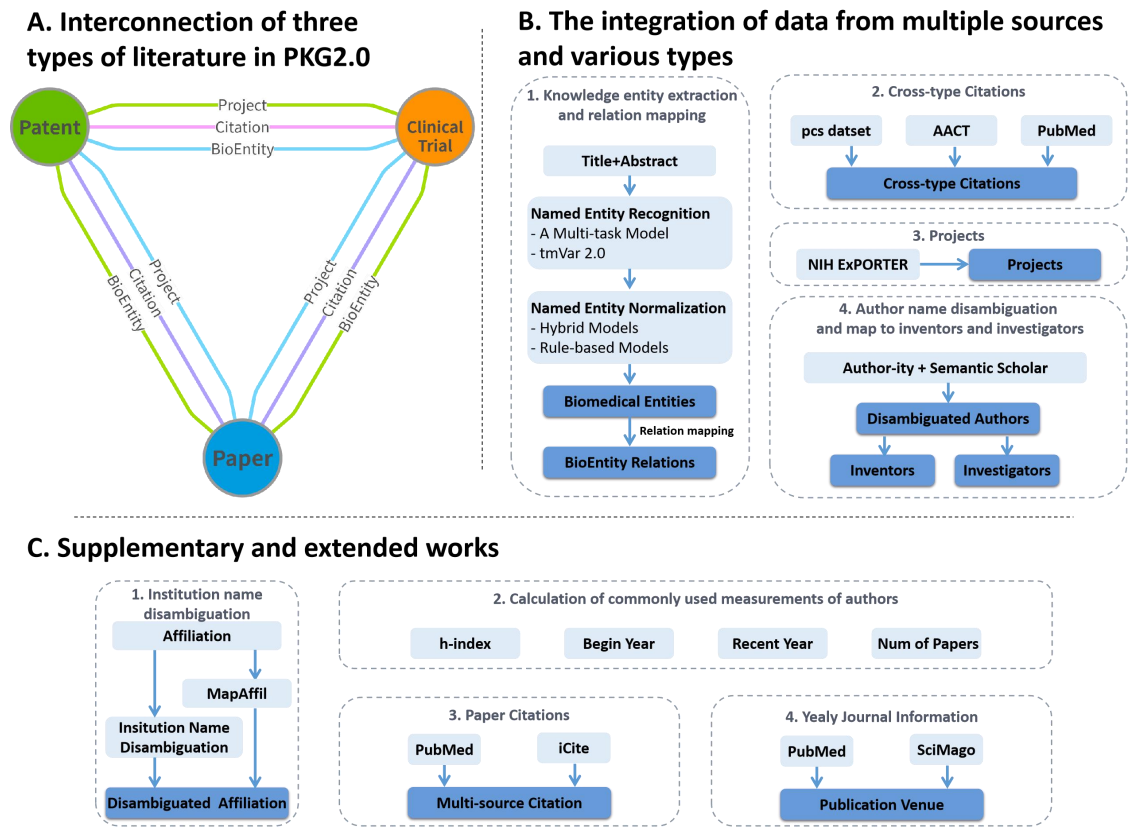

Fig. 1 Framework of PKG 2.0. **(A)** Three ways in which the three types of academic literature link to each other. **(B)** The specific methods or data sources of the linkage. **(C)** Supplementary and extended works in PKG 2.0.

The core of PKG includes the entirety of PubMed papers, all clinical trial studies from ClinicalTrials.gov (https://clinicaltrials.gov/), and biomedical-related patents from USPTO (https://www.uspto.gov/) and PatentsView (https://patentsview.org/). PKG establishes connections between papers, patents, and clinical trials through bioentities, projects, authors, and citations. It integrates extended metadata from multiple sources, including iCite (https://icite.od.nih.gov/), and SciMago (https://www.scimagojr.com/). This results in a biomedical knowledge graph that not only features rich metadata but also provides fine-grained knowledge annotations and tightly interconnected relationships. Figure 2 shows the logical model of PKG in a relational database.

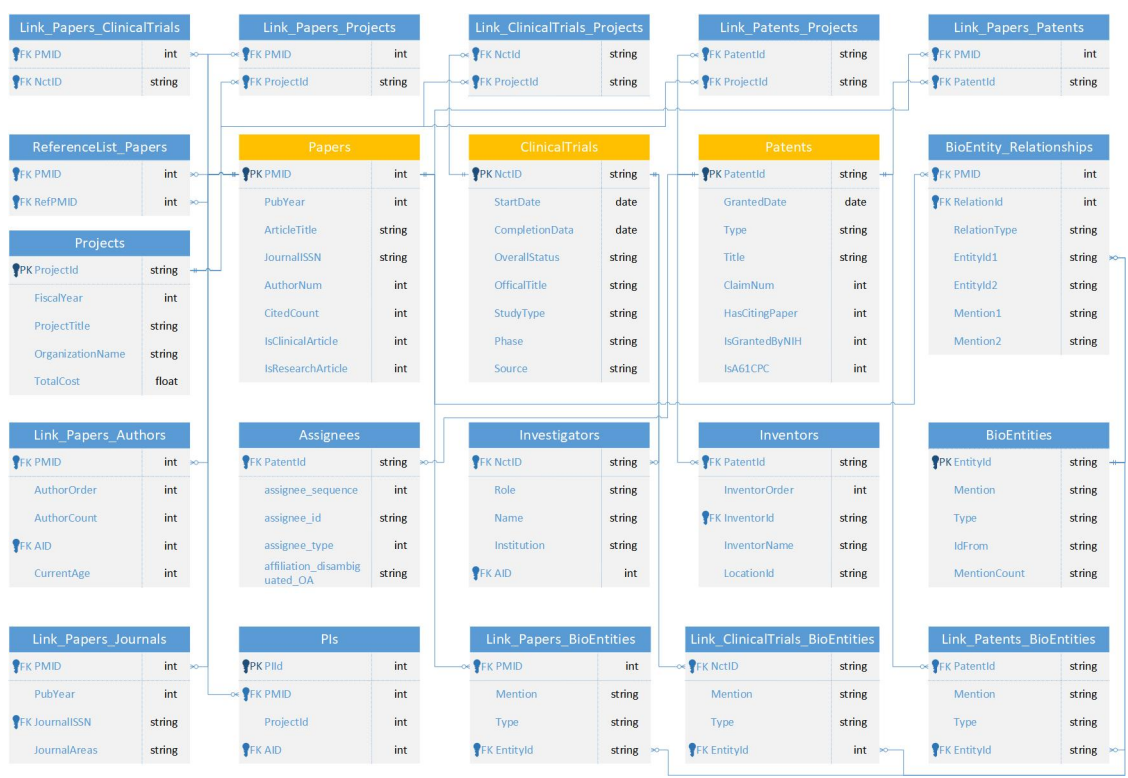

Fig. 2 The entity relationship diagram of PKG. PKG includes "Articles", "ClinicalTrials", and "Patents" as the main data tables, and links them to each other through a series of relationship tables, as well as to other tables capturing data from a range of sources. For clarity, here we show a subset and their main fields of the tables (see the Data Records section and the appendix for a more comprehensive view of the tables).

In practice, PKG 2.0 is expected to benefit a diverse range of users across multiple fields. For those in the information retrieval domain, researchers can access a wealth of related information from different types of literature, which significantly enhances the comprehensiveness and accuracy of knowledge retrieval compared to traditional search methods. In the field of scientometrics, the integrated nature of PKG 2.0 allows scholars to analyze the relationships between different types of scholarly outputs, such as citation patterns, collaboration networks, and knowledge transfer mechanisms among papers, patents, and clinical trials. This holistic view can uncover previously hidden patterns and trends, facilitating more profound scientometric analysis and research. In the realm of machine learning and artificial intelligence, PKG 2.0 offers high-quality linked datasets that are highly conducive to knowledge reasoning. The well-structured and interrelated data can be used to train models, enabling machines to better understand complex relationships within biomedical research and ultimately improving the performance and accuracy of machine-learning algorithms in tasks such as prediction, classification, and decision-making in the biomedical field.

## Methods

PKG encompasses the entirety of PubMed papers, all clinical trial studies from ClinicalTrials.gov, and biomedical-related patents from USPTO. Linkages between these three types of literature are established through knowledge entities, citations, author name disambiguation (AND), and projects provided by NIH Exporter (https://reporter.nih.gov). Fine-grained connections at the knowledge entity level can provide more accurate insights into the flow of knowledge. Subsequent work includes matching journal information and disambiguating institutions. These efforts rely on high-quality fine-grained entities, comprehensive citation data, accurate author disambiguation, and reliable project information. Therefore, the extraction of knowledge entities, author disambiguation, citation integration, and project details across papers, clinical trials, and patents are crucial components of this study.

## Linkages based on fine-grained knowledge entities

In this section, we detail the methodologies employed to establish linkages between papers, clinical trials, and patents using fine-grained knowledge entities. The unique identifiers assigned to these knowledge entities serve as a primary mechanism for interconnecting the three types of literature. This approach is crucial for creating a comprehensive and interconnected knowledge graph that accurately reflects the complex relationships within biomedical research. The primary challenge lies in the precise extraction and normalization of bio-entities, ensuring that each entity is uniquely and consistently identified across different document types. Additionally, mapping the relationships between these entities requires sophisticated techniques to handle the vast and diverse datasets involved. To address these challenges, we employ advanced natural language processing methods and leverage the iBKH dataset (https://github.com/wcm-wanglab/iBKH/)[57] for accurate relation mapping. This section will cover the detailed processes of biomedical entity extraction, normalization, and relation mapping, as well as the extraction of datasets and methods entities.

### Biomedical entity extraction and relation mapping

Knowledge entities provide a quick understanding of the research content or topics within the literature, facilitating further literature management and pattern analysis. Interactions between genes, proteins, or drugs in specific biological contexts are central to many biomedical studies. However, the vast volume and rapid publication rate of biomedical literature make manual identification of these content elements impractical[58]. To address this challenge, named entity recognition techniques are employed to automatically identify domain-specific terms in PubMed articles.

We propose a biomedical text mining tool, BERN2 (http://bern2.korea.ac.kr/)[59], for accurate and efficient biomedical Named Entity Recognition (NER) and named entity normalization (NEN). BERN2 has been used to extract biomedical entities from the titles and abstracts of papers, clinical trials, and patents in this study. The extracted entities include nine categories: gene/protein, disease, drug/chemical, species, mutation, cell line, cell type, DNA, and RNA. The bio-entity extraction component comprises two models: a multi-task NER model that identifies named entities in papers, clinical trials, and patents, and a multi-type normalization model that links these biomedical entities to unique concept identifiers. After performing biomedical entity extraction, we used the iBKH dataset to map the relationships between the extracted entities.

(1) Named entity recognition. BERN2 employs a multi-task NER model to simultaneously extract all entity types except for mutations. Multi-task learning (MTL)[60] aims to improve performance by allowing models to share representations across related tasks.

In biomedical NER, MTL often outperforms single-task learning (STL) because it enables models to leverage shared word semantics across different tasks[61]. Due to the lack of a publicly available training set specifically for mutations, BERN2 utilizes tmVar 2.0[62] as the NER model for mutation-type biomedical entities. Cross-entropy objectives were used to optimize the model, and the loss function for the model was defined as follows:

$$\mathcal{L} = -\frac{1}{NM}\sum_{i=1}^{N}\sum_{j=1}^{M} y_{ij} \cdot \log(p_{ij}) \qquad (1)$$

where $N$ is the number of task-specific layers, $M$ is the max sequence length of input texts, $y_{ij}$ denotes the ground-truth label, and $p_{ij}$ denotes the probability that each task-specific layer produces.

(2) Named entity normalization. Relying solely on rule-based NEN models cannot cover all variations in entity names. To address this limitation, we use the BioSyn[63] module, a neural network-based NEN approach, which extends standardization to entities not covered by rule-based methods. Regarding type overlaps, such as the case of 'androgen' belonging to both genes and compounds, we prioritize categories with CUIs, as they provide richer information. Additionally, mutations identified by tmVar2.0 consistently receive the highest priority category label.

(3) Using the iBKH dataset to label relations between bio-entities. The study of relationships between entities holds great promise for advancing our understanding of complex biological systems and improving healthcare outcomes. It can enhance the predictive capabilities of biomedical models, leading to more accurate diagnoses and more effective treatments. Based on the aforementioned BERN2 entity extraction results, we combined the iBKH to map the relationships between biomedical entities. Specifically, a total of six types of relationships were mapped, including disease-drug, disease-disease, disease-gene, drug-drug, drug-gene, and gene-gene. As BERN2 and iBKH use different sources of entity IDs, we mapped different IDs during the mapping process. BioMart[64], Human Disease Ontology[65], PharmGKB[66], and drug_id_mapping[67] were used to map the entity IDs in iBKH to the entity IDs in BERN2.

**Dataset and method entity extraction**

Identifying the methods and datasets used in academic papers is crucial for ensuring reproducibility, validating results, and enabling further research advancements.

We used a dictionary-based approach to extract methods and datasets. Relative words in Ontology for Biomedical Investigations[68] are used for method entity extraction, and NAR Database (https://www.oxfordjournals.org/nar/database), NIH-BMIC's domain-specific repositories (https://www.nlm.nih.gov/NIHbmic), and EMBL-EBI data resources (https://www.ebi.ac.uk/services/data-resources-and-tools) are used for data set entity extraction. This is an exploratory fine-grained entity extraction work, and the problem of name ambiguity still needs to be resolved.

## Linkages based on citations

In this section, we outline the methodologies employed to link papers, patents, and clinical trials through citation relationships. Citations serve as one of the primary mechanisms for interconnecting these three types of literature, providing a comprehensive view of how knowledge flows and evolves across different domains. This approach is essential for creating a cohesive and interconnected knowledge graph that accurately reflects the citation networks within biomedical research. The primary challenge lies in integrating citation data from multiple sources and ensuring the accuracy and completeness of these linkages. To address these challenges, we leverage various datasets and identify specific citation patterns to enhance the reliability of citation data. This section will cover the detailed processes of integrating citations from multiple sources and establishing citation linkages among papers, patents, and clinical trials.

**Integrate citations from multiple sources**

Complete citation data is fundamental for ensuring the reliability of citation analyses[69]. Thanks to the open citation movement, the number of citations in academic literature has surpassed one billion, and the reliability of open citation data now exceeds that of closed databases[70]. To enhance the completeness and reliability of citation data, we integrated data from two sources:

(1) PubMed Citation Data: PubMed citation data is provided by various publishers. However, the level of attention given to this task varies among publishers, resulting in significant data gaps.

(2) NIH Open Citation Collection (NIH-OCC)[71]: NIH-OCC data consists of PMID-to-PMID citation relationships. The non-structured citation text from articles is initially extracted and matched with relevant articles, alongside details on the fields present in the paper (e.g., journal name, author names, title terms). The comparison of fields such as title, author, and journal name with the input text helps in pinpointing the most suitable matches. NIH-OCC data is generated from unrestricted sources such as MedLine, PubMed Central (PMC), and CrossRef, resulting in higher data completeness.

During the integration process, through manual verification, we found that some citation data might have incorrect links. For example, some data from earlier years in PubMed contain obvious errors, especially where the publication year of the citing literature is several years later than that of the cited literature. To address this, we removed citation data with a negative difference of more than 1 year.

**Citations between papers and patents**

Technological innovation serves as the foundation for economic growth, and many valuable innovations rely on scientific discoveries[72]. In fact, science can be considered a 'map' for business inventors who seek to leverage original technologies or ignite their own research and development efforts[73]. The citations between scientific papers reflect mutual influences among research endeavors. Additionally, the direct citation of patents in academic papers (patent-to-paper citations) provides a supplementary measure of the impact of fundamental research on commercialization. However, extracting patent-to-paper citations poses challenges. Non-patent literature citations recorded by the United States Patent and Trademark Office (USPTO) do not always adhere to standard citation formats, and a significant number of citations exist only within the main text, not in the patent metadata. Consequently, systematically identifying referenced articles remains a complex task.

In PKG, we established links between PubMed articles and USPTO patents through citations. We used the V64 version of the PCS dataset

(https://zenodo.org/records/11461587/)[74] as the base data. By mapping Digital Object Identifiers (DOIs) to PubMed Identifiers (PMIDs), we were able to filter and identify USPTO patent citations that were specifically related to PubMed articles. Through this meticulous process, we compiled a comprehensive dataset comprising a total of 25,597,962 rows of data. This dataset encompasses 2,400,855 unique USPTO patents and 3,112,470 distinct PubMed articles. The integration of these two significant data sources provides a robust framework for analyzing the interplay between scientific research and technological innovation, facilitating a deeper understanding of the citation dynamics between academic publications and patent literature.

**Citations between papers and clinical trials**

Journal articles tend to report the results of experiments, while detailed experimental procedures and additional information are often documented in clinical trial reports. In the United States, primary trial researchers funded by the National Institutes of Health (NIH) are actively encouraged to publish trial results in journal articles[75]. The reporting of trial results is accomplished through publishing journal articles containing trial results and submitting summaries to trial registries. Ideally, each trial should have at least one article about the results, even if the trial outcomes are non-significant or negative. However, scientists often tend to publish only their positive findings, leading to a biased exploration of trial progress based solely on published papers. This approach may even result in significant errors in conclusions.

Analyzing clinical trial registrations fundamentally helps mitigate this bias[76]. In PKG, we addressed this publication bias issue by incorporating Clinical Trial data. The Clinical Trial dataset is sourced from ClinicalTrials.gov and provides information to the public, researchers, and healthcare professionals regarding clinical studies and their outcomes. Aggregate Analysis of ClinicalTrials.gov (AACT) Database (https://aact.ctti-clinicaltrials.org/)[77] processed ClinicalTrials.gov data, organizing it into 54 relational data tables that are updated daily. We selected references related to PubMed articles from the AACT dataset. In addition, the Uniform Requirements for Manuscripts (URM)[78] requires timely trial registration before recruiting the first trial participant and ensuring accurate linkage between published journal articles and trial registration records. PubMed includes trial registration numbers in its provided XML files, and we processed this data to as links between clinical trials and papers in PKG.

By incorporating clinical trial reports into PKG, we can trace the origins of research outcomes in papers and also retrospectively explore the academic theoretical foundations of clinical experiments, thus fostering cyclical advancement in biomedical scientific research.

**Citations between patents and clinical trials**

We have established links between patents and clinical literature through citations. Specifically, if a patent citation contains the character "CT" followed by an 8-digit number, or if a clinical trial citation includes "us" or "patent" followed by an 8-digit number, the corresponding citation is extracted. It is important to note that the number of citation links between patents and clinical literature is relatively small.

# Linkages based on projects

The NIH Exporter serves as a valuable resource, offering a wealth of information about NIH-funded projects. This includes not only the metadata of the projects themselves but also the associated scholarly outputs in the form of papers, patents, and clinical trials. The annual publication of the latest data ensures that the information remains current and relevant, providing a comprehensive overview of the landscape of NIH-funded research. According to our investigation, NIH-funded research accounts for a significant proportion of all grants recorded in PubMed.

By linking the metadata of NIH-funded projects with the corresponding papers, patents, and clinical trials, we establish a detailed and interconnected view of the research landscape. This integration not only enriches the data within PKG but also enhances its analytical capabilities. It allows for a deeper understanding of the relationships and impacts of NIH-funded research across different types of literature. This comprehensive approach is crucial for facilitating advanced analyses and uncovering insights that were previously hidden due to the fragmented nature of scientific literature.

**Supplementary and extended works**

In this section, we outline the supplementary and extended works that enhance the functionality and comprehensiveness of PKG. This section will cover the processes of supplementing journal information by year of publication, institution name disambiguation, and calculation of commonly used measurements.

**Author name disambiguation**

The issue of name ambiguity in scholarly publications is one of the most complex and highly scrutinized aspects of disambiguation. Based on our investigation, by integrating existing AND datasets such as Author-ity (https://databank.illinois.edu/datasets/IDB-2273402/)[79] and Semantic Scholar (https://www.semanticscholar.org/)[80], a comprehensive and high-quality disambiguation result for PubMed author names can be obtained.

We selected the Author-ity dataset's author disambiguation results as the primary author identify (AND_ID) and supplemented it with Semantic Scholar's AND results for author instances after 2018. Due to variations in disambiguation results from different sources, a single AND_ID in Semantic Scholar may simultaneously map to multiple AND_IDs in the Author-ity dataset. This occurs due to Author-ity assigning different AND_IDs to the same author or AND_IDs being incorrectly assigned by Semantic Scholar. To address this propagation of errors during the integration process, we additionally trained a deep neural network (DNN) model to merge conflicting AND_IDs using information such as author names, institutions, emails, countries, additional name information from OpenAlex, and title and abstract embeddings, etc. The entire author disambiguation process can be outlined in the following steps, as shown in Figure 3:

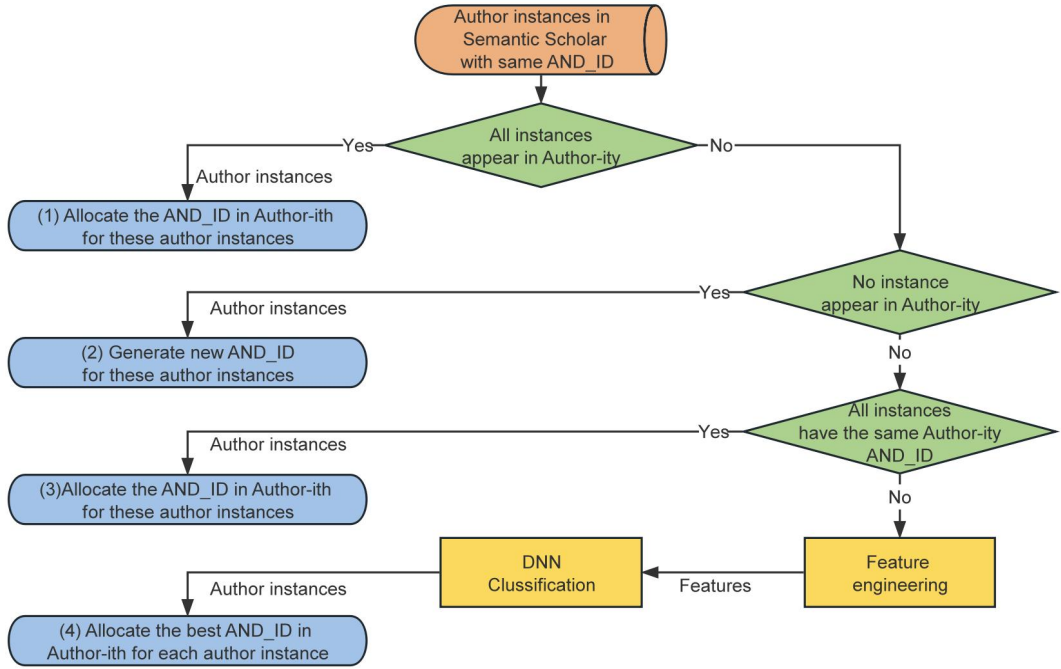

Fig. 3 The process of the author name disambiguation. It mainly combines the author disambiguation results of the Author-ity dataset and Semantic Scholar.

As illustrated in Figure 3, the author name disambiguation process in PKG involves the following steps:

(1) We assigned a unique ID to each author instance based on the Author-ity AND results, ensuring that authors from the Author-ity dataset (before 2018) have distinct IDs. This guarantees the highest accuracy for author instances within the Author-ity dataset.

(2) For authors with the same Semantic Scholar AND_ID who never appeared in the Author-ity dataset, we created a new AND_ID to label them. For instance, author "Florian Bucher" published two papers in 2021, resulting in two corresponding author instances. Because all papers that "Florian Bucher" published were after 2018, they were not included in the Author-ity dataset and thus had no AND_ID assigned by Author-ity. However, authors correctly disambiguated by Semantic Scholar were given unique AND_IDs in Semantic Scholar. To maintain consistent labeling, we generated a new AND_ID, continuing from the Author-ity AND_IDs, to label these two author instances as disambiguated by Semantic Scholar.

(3) For author instances with a unique AND_ID in Semantic Scholar that also have the same AND_ID in the Author-ity dataset, we assigned the Author-ity AND_ID to all instances as their unique ID. For example, "Thomas Hofbeck" published nine papers between 2010 and 2021. The first eight instances had a unique Author-ity AND_ID, but the last one did not, as it was beyond the Author-ity dataset's time coverage. However, based on the AND results from Semantic Scholar, all nine instances had the same AND_ID. Therefore, the last instance without an Author-ity AND_ID was labeled with the same ID as the other eight instances.

(4) For authors instances with a unique AND_ID in Semantic Scholar but have different AND_IDs in the Author-ity dataset, we trained a DNN model to handle this conflict to allocate a best AND_ID. The model treats author instances with the same AND_ID as author groups and determines the most likely author group for the conflicting author instance. We trained the DNN model using the LAGOS-AND dataset (https://zenodo.org/records/6731767/)[81], creating approximately a 1:1 ratio of positive and negative samples. The model features include author names, institutions, emails, countries, additional name information from OpenAlex, and title and abstract embeddings. The validation results show that PKG significantly improves recall and F1 scores while maintaining precision.

Additionally, we utilized this trained model to conduct name disambiguation for patent inventors and clinical trial investigators. By applying the similar set of features and the learned disambiguation model, we aim to identify and disambiguate individuals in these literatures. It should be noted that since the AND_ID in PKG is the result of a probabilistic learning model, the disambiguation result is not recommended for individual research.

**Institution name disambiguation**

Unified names and identifiers are essential for ensuring the reliability of institution-related studies. However, there is a lack of institution name disambiguation in the affiliations of paper authors, clinical trial investigators, and patent assignees in general. Performing institution disambiguation ensures that researchers accurately refer to the same institutions when analyzing and comparing data from papers, clinical trials, and patents.

While institution name repetition is not as frequent as that of author names, it remains a challenging task. The first hurdle lies in extracting institutional information because author affiliations are often unstructured. A genuine affiliation may encompass various elements, including country, state, city, institution, and department. These affiliations are influenced by individual author's habits and publisher formatting styles. To gain a better understanding of author affiliations, we employed the MapAffil[82,83] for affiliation extraction. This tool automatically parses information from free-text affiliation fields, extracting details such as country, state, city, location, institution, and department.

Despite the MapAffil effectively captures various aspects of authorship at the individual level, further institution disambiguation is necessary for analyzing institutions at a higher level (e.g., institutional rankings). Different scholars may employ distinct writing styles and coding conventions when referring to the same institution[84]. For instance, variations like "univ instelling antwerp," "univ antwerpen ua," "univ antwerpen cde," "univ antwerp vib," and "univ antwerp uia" all correspond to "univ antwerp." Additionally, due to translation errors and misspellings, institutional ambiguity is prevalent in academic literature datasets[85]. Without institution name disambiguation, institutions with irregularly written names may appear weaker in rankings compared to equally capable counterparts. Therefore, instituting further disambiguation processes is crucial for accurate and consistent representation of institutions in scholarly assessments. We used OpenAlex's solution to solve this problem. OpenAlex addresses this issue by combining two independent classification models[86]. These models were trained on extensive historical MAG[87] data and ROR[88] data. According to the model's report[86], its precision is 0.96, and its recall is 0.95.

We implemented this institution disambiguation work on the affiliations of paper authors, clinical trial investigators, and patent assignees. This application to these three types of literature data-related affiliations helps to ensure the accuracy and consistency of institutional representation in various academic and research contexts.

**Supplementing specific information of the journal by year of publication.**

In the assessment of scientific capabilities for research institutions and individuals, the primary focus is on the achievements of publicly published papers. The relative quality of the journals where these papers are published serves as a direct, crucial, and popular evaluation criterion. For complex scientific research questions that cannot be adequately addressed within a single field, interdisciplinary research often provides more effective solutions[89]. Consequently, understanding the quality of journals in other fields has become increasingly important.

PubMed provides information on the source journals for each indexed article. However, PubMed encompasses over thirty thousand different journals, and even seasoned scholars may struggle to recall the quality levels of journals across various domains. Therefore, quantitative evaluation of journals becomes essential. SCImago is a publicly accessible portal that provides journal metrics calculated based on information from the Scopus database. These metrics serve for evaluating and analyzing scientific fields. We linked SCImago data to PKG through the following steps:

(1) Completing ISSN Fields in PubMed Journals: The ISSN (International Standard Serial Number) uniquely identifies journals and facilitates convenient and accurate cross-database linking. Approximately 1% of journals in PubMed lack an ISSN field. To address this, we supplemented the ISSN data using other journal names as references.

(2) Linking PKG with SCImago Using ISSN and PubYear Fields: We believe that evaluating paper quality based on journal information should not be static. Instead, it should consider the journal's status in the year of publication. Rankings of different journals within the same field can fluctuate significantly across years. When specific journal information is unavailable for the publication year, we use data from the nearest adjacent year.

(3) Supplementing Data Based on the Linked Results: Using the results from the second step, we added data for the publication year, including SJR (Scientific Journal Rankings), impact factor, h-index, total publications, journal categories, publishers, and geographic regions.

**Calculation of commonly used measurements**

Understanding the contributions and impact of individual researchers is crucial for evaluating scientific progress and identifying areas of disparity. By accurately disambiguating authors and calculating key scientific metrics, we can gain deeper insights into the academic landscape. This work not only helps in recognizing the achievements of individual researchers but also sheds light on broader trends and patterns within the scientific community. Based on the author disambiguation results, we calculated scientific metrics for each author. Specifically:

(1) Counting of the author's publications: This is based solely on articles indexed in PubMed. If an author's articles are not included in PubMed, the calculated results may be lower than the actual count. However, considering PKG's focus on biomedical literature, we find this approach reasonable.

(2) Calculation of the h-index: This is a comprehensive metric that evaluates an author's scholarly impact by considering both the number of publications and the number of citations received. It is widely used to assess academic capabilities.

# Data Records

The dataset is available at Figshare[90], and keep update on GitHub at https://PubMedKG.github.io. The statistics and descriptions of the main nodes and relationships in PKG are presented in Tables 2 and 3, respectively. See the "PKG 2.0 Database Description.pdf" at Figshare[90] for more details of PKG. To help users organize and utilize the dataset more effectively, we have added serial numbers to the table names in the file available on Figshare.

Table 2. Statistics of node types

| Node type | Statistics | Describe |
| --- | --- | --- |
| Article | 36,551,113 | Papers from PubMed |
| Patent | 1,344,469 | Patents from USPTO |
| Clinical Trials | 480,795 | Clinical trial studies from ClinicalTrials.gov |
| Author | 26,217,594 | Disambiguated authors in PKG |
| Institution | 69,457 | Disambiguated institutions identified with ROR |
| BioEntity | 357,686 | BioEntities extracted by BERN2 |
| Journal | 21,382 | Journals mapped with SciMago |
| Project | 2,023,148 | Projects funded by NIH |

Table 3. Statistics of relation types

| Relation type | Node type 1 | Node type 2 | # of Relation |
| --- | --- | --- | --- |
| Paper-Author | Paper | Author | 160,848,959 |
| Paper-Journal | Paper | Journal | 29,215,730 |
| Paper-Paper (citation) | Paper | Paper | 774,810,780 |
| BioEntity-Clinical Trial | BioEntity | Clinical Trial | 11,416,614 |
| BioEntity-Paper | BioEntity | Paper | 464,643,559 |
| BioEntity-Patent | BioEntity | Patent | 6,469,587 |
| Project-Clinical Trial | Project | Clinical Trial | 33,658 |
| Project-Paper | Project | Paper | 6,984,399 |
| Project-Patent | Project | Patent | 81,133 |
| Author-PI | Author | PI | 53,424,596 |
| Paper-Clinical Trial (citation) | Paper | Clinical Trial | 967,719 |
| Paper-Patent (citation) | Paper | Patent | 18,192,186 |

We performed biomedical entity extraction on articles, clinical trials, and patents. Table 4 presents statistical data on nine types of entities extracted from the three types of literature. The extraction results for each of the three types of literature are stored in separate data tables.

Table 4 Statistics of extracted biomedical entities

| | Article | | Clinical Trial | | Patent | |
| --- | --- | --- | --- | --- | --- | --- |
| Bio-entity Type | Total number | Distinct PMIDs | Total number | Distinct nct_ids | Total number | Distinct PatentIds |
| Species | 46,363,597 | 11,610,019 | 2,097,318 | 361,618 | 501,642 | 173,495 |
| Disease | 101,605,033 | 17,767,719 | 3,063,362 | 366,880 | 606,955 | 192,240 |
| Gene | 82,177,926 | 9,444,042 | 703,203 | 125,248 | 703,600 | 193,360 |
| Drug | 89,639,755 | 12,542,164 | 1,588,577 | 226,824 | 2,178,980 | 371,800 |
| mutation | 1,018,182 | 263,564 | 5,576 | 2,250 | 5,308 | 2,307 |
| cell_line | 13,616,107 | 5,090,803 | 56,917 | 28,114 | 181,511 | 97,499 |
| cell_type | 26,530,138 | 8,252,231 | 294,149 | 98,660 | 210,375 | 103,579 |
| DNA | 22,920,173 | 7,660,784 | 165,055 | 80,523 | 787,810 | 324,921 |
| RNA | 4,436,313 | 1,809,388 | 20,063 | 10,212 | 5,308 | 2,307 |

Each data field is self-explanatory, and fields with the same names in other tables adhere to a consistent data format that allows cross-table linking. Tables 5 to 24 provide details on the essential field names, formats, non-empty data counts, and brief descriptions for each data file.

Table 5 Data type for records of Papers

| Index | Format | Short description |
|---|---|---|
| PMID | Integer | Unique ID assigned by PubMed to identify articles. |
| PubYear | Integer | The year in which the journal issue was published. |
| ArticleTitle | String | The title of the article. |
| AuthorNum | Integer | The number of the Author. |
| CitedCount | Integer | The number of citations. |
| IsClinical | Integer | Mark whether this article belongs to a clinical trial. |

Table 6 Data type for records of Link_Papers_Authors

| Index | Format | Short description |
|---|---|---|
| PMID | Integer | Unique ID assigned by PubMed to identify articles. |
| AuthorOrder | Integer | The order of the author in this article. |
| AuthorNum | Integer | The number of authors in this article. |
| PubYear | Integer | The year in which the journal issue was published. |
| AID | Integer | Author ID in PKG. |

Table 7 Data type for records of Affiliations

| Index | Format | Short description |
|---|---|---|
| PMID | Integer | Unique ID assigned by PubMed to identify articles. |
| AuthorOrder | Integer | The order of the author in this article. |
| AID | Integer | Author ID in PKG. |
| Affiliation | String | Raw affiliation string in Pubmed. |
| Department | String | Department of the affiliation. |
| Institution | String | Institution of the affiliation. |
| Zipcode | String | Zipcode of the affiliation. |
| Country | String | Country of the affiliation. |
| City | String | City of the affiliation. |
| State | String | State of the affiliation. |
| Type | String | Type of the institution. |
| IND_ID | Integer | Unique ID of the institution. |

| Institution_IND | String | Institution of the author after the institution name disambiguation. |

Table 8 Data type for records of ReferenceList_Papers

| Index | Format | Short description |
| --- | --- | --- |
| PMID | Integer | The PMID of the citing article. |
| RefArticleID | Integer | The PMID of the cited article. |

Table 9 Data type for records of PIs

| Index | Format | Short description |
| --- | --- | --- |
| PI_ID | Integer | A unique identifier for each of the project Principal Investigators. |
| PMID | Integer | Unique ID assigned by PubMed to identify articles. |
| AID | Integer | Author disambiguation ID. |
| Application_ID | String | A unique identifier of the project record in the ExPORTER database. |

Table 10 Data type for records of Link_Papers_BioEntities

| Index | Format | Short description |
| --- | --- | --- |
| PMID | Integer | Unique ID assigned by PubMed to identify articles. |
| StartPosition | Integer | Start position of mention in an abstract. |
| EndPosition | Integer | End position of mention in an abstract. |
| Mention | String | Entity mentioned in an abstract. |
| Entityid | String | Normalized entity identifiers, include mesh, mim, CL, cellosaurus, NCBITaxon, NCBIGene, CHEBI. |
| Type | Integer | Enumerated type of entity; values include species, disease, gene, drug, mutation, cell_line, cell_type, DNA, RNA. |
| is_neural_normalized | Integer | For diseases and chemicals, BERN2 use hybrid NEN models, which are a combination of both rule-based and neural network-based models. An entity that is not normalized by the rule-based model is then normalized by a neural network-based model. |

Table 11 Data type for records of Link_Papers_Journals

| Index | Format | Short description |
| --- | --- | --- |
| PMID | Integer | Unique ID assigned by PubMed to identify articles. |
| PubYear | Integer | The year in which the journal issue was published. |
| Journal_ISSN | String | Unique ID of Journal. |
| Journal_Title | String | Title of Journal. |
| Journal_SJR | Float | The SJR of the journal when the issue was published. |
| Journal_Hindex | Integer | The h-index of the journal when the issue was published. |

Table 12 Data type for records of ClinicalTrials

| Index | Format | Short description |
| --- | --- | --- |
| nct_id | String | Unique ID assigned by ClinicalTrials.gov to identify clinical trial studies. |
| brief_title | String | Title of the trial. |
| start_date | date | The date the study started in a date-type format so that it can be used to find studies before/after/inclusive of a date or dates. For studies that only provide |

| | | month/year, the last day of the month is used. |

Table 13 Data type for records of Link_Papers_ClinicalTrials

| Index | Format | Short description |
|---|---|---|
| PMID | Integer | Unique ID assigned by PubMed to identify articles. |
| NctId | String | Unique clinical trial study identifier in ClinicalTrials.gov. |

Table 14 Data type for records of Link_ClinicalTrials_BioEntities

| Index | Format | Short description |
|---|---|---|
| nct_id | String | Unique ID assigned by ClinicalTrials.gov to identify clinical trial studies. |
| Entityid | String | Normalized entity identifiers, include mesh, mim, CL, cellosaurus, NCBITaxon, NCBIGene, CHEBI. |

Table 15 Data type for records of Investigators

| Index | Format | Short description |
|---|---|---|
| nct_id | String | Unique ID assigned by ClinicalTrials.gov to identify clinical trial studies. |
| AID | Integer | Unique ID assigned by PKG to identify authors. |
| same_author_prob | Float | The score of the linkage between the investigator and the AID. |

Table 16 Data type for records of Patents

| Index | Format | Short description |
|---|---|---|
| PatentId | String | Unique ID assigned by USPTO to identify patents. |
| GrantedDate | Date | Granted date of the patent. |
| Title | String | Title of the patent. |
| Abstract | String | Abstract of the patent. |

Table 17 Data type for records of Link_Patents_Papers

| Index | Format | Short description |
|---|---|---|
| PatentId | String | Unique ID of patent in USPTO. |
| RefPMID | Integer | Unique ID assigned by PubMed to identify articles. |
| ConfScore | Integer | Confidence of the citation. |
| WhereFound | String | Where the citation was been founded. |

Table 18 Data type for records of Assignee

| Index | Format | Description |
|---|---|---|
| patent_id | String | Unique ID assigned by USPTO to identify patents. |
| assignee_sequence | Integer | Order in which assignee appears on the patent. |
| assignee_id | String | Unique PatentsView database assignee ID assigned by disambiguation algorithm. |
| disambig_assignee_organization | String | Organization name, if assignee is organization. |
| assignee_type | Integer | Classification of assignee:2 - US Company or Corporation, 3 - Foreign Company or Corporation, 4 - US Individual, 5 - Foreign Individual, 6 - US Government, 7 - Foreign Government, 8 - Country Government, 9 - State |

| | | Government (US). Note: A "1" appearing before any of these codes signifies part interest. |
| --- | --- | --- |
| affiliation_disambiguated_OA | String | Unique ID and organization name assigned by OpenAlex's disambiguation algorithm. |

Table 19 Data type for records of Link_Patents_BioEntities

| Index | Format | Short description |
| --- | --- | --- |
| PatentId | String | Unique ID of patent in USPTO. |
| StartPosition | Integer | Start position of mention in an abstract. |
| EndPosition | Integer | End position of mention in an abstract. |
| Mention | String | Entity mentioned in an abstract. |
| Entityid | String | Normalized entity identifiers, include mesh, mim, CL, cellosaurus, NCBITaxon, NCBIGene, CHEBI. |
| Type | Integer | Enumerated type of entity; values include species, disease, gene, drug, mutation, cell_line, cell_type, DNA, RNA. |

Table 20 Data type for records of Inventors

| Index | Format | Short description |
| --- | --- | --- |
| PatentId | String | Unique ID of patent in USPTO. |
| inventor_id | String | Inventor id assigned by PatentsView. |

Table 21 Data type for records of BioEntity_Relationships

| Index | Format | Short description |
| --- | --- | --- |
| PMID | Integer | Unique ID assigned by PubMed to identify articles. |
| entity_id1/entity_id2 | String | The IDs of two biomedical entities. |
| relation_type | String | Type of the two biomedical entities. |
| relation_id | String | ID allocated by PKG to identify the relation. |

Table 22 Data type for records of DatasetMethod

| Index | Format | Short description |
| --- | --- | --- |
| PMID | Integer | Unique ID assigned by PubMed to identify articles. |
| Mention | String | Entity mentioned in an abstract. |
| Entity | String | Standard expression of entity. |
| Type | Integer | Identify whether this entity is a method or a dataset. |

Table 23 Data type for records of BioEntities

| Index | Format | Short description |
| --- | --- | --- |
| EntityId | String | Normalized entity identifier. |
| Type | String | Enumerated type of entity; values include species, disease, gene, drug, mutation, cell_line, cell_type, DNA, RNA. |
| Mention | String | Text of the bioentity. |

Table 24 Data type for records of Link_Clinicaltrials_Patents

| Index | Format | Short description |
| --- | --- | --- |

| nct_id | String | Unique ID assigned by ClinicalTrials.gov to identify clinical trial studies. |
| PatentId | String | Unique ID of patent in USPTO. |
| Type | String | Type of the citation; values include citing and cited. |

# Technical Validation

## Validation of author name disambiguation

We validate our disambiguation results using the LAGOS-AND dataset. This dataset is a recently published high-quality collection of author name disambiguation data, automatically constructed using authoritative sources such as ORCID and DOI. According to its reported validation information, the LAGOS-AND dataset is highly credible. The 2.0 version of the LAGOS-AND dataset is built on the OpenAlex database's April 2022 baseline version.

Table 25 The comparison of the accuracy of AND

|  | % Precision | % Recall | % F1 score |
| --- | --- | --- | --- |
| Author-ity | 98.58 | 67.79 | 80.33 |
| Semantic Scholar | 98.53 | 86.55 | 92.15 |
| PKG | 98.45 | 94.10 | 96.24 |

According to the evaluation results presented in Table 25, the Author-ity dataset exhibits the highest precision. Our integrated results ensure credible precision while significantly improving the recall and F1 score.

## Validation of citations

The combined data table "ReferenceList_Papers" contains 774,810,780 citation relationships across 30,695,697 articles. Compared to PubMed's native citation data (268,400,936 entries), the cumulative supplementation amounts to 506,409,844 additional citations. Validations of cross-type citations are as follows:

### Validation of paper-patent citations

Matt[74] provided a high-quality PCS dataset. We selected a portion of the data from USPTO to PubMed for validation in our dataset. This dataset contains only positive annotations and covers citations from 183 patents, which can be used to calculate recall. However, to evaluate precision, we need negative annotations. Therefore, we supplemented the labeled data for these 183 patents to compute accuracy. For each confidence level, we randomly extracted 100 annotation samples. During the annotation process, we primarily used auxiliary information such as author names, publication year, article title, journal, issue, and volume. Each annotated data point underwent verification by two research assistants.

Based on the known good dataset provided by Matt[74], we extracted the relevant portion related to PubMed (4,223 rows) and calculated the recall of the PCS dataset in PKG at different confidence levels.

Table 26 The accuracy of patent-paper linkage

| Confidence | Non-patent references linked | Manually marked Incorrect | % Precision | Estimated cumulative % Precision | # found (of 4,223 known-good) | % recall |
| --- | --- | --- | --- | --- | --- | --- |

|  |  | （sample of 100） |  |  |  |  |
|---|---|---|---|---|---|---|
| 10 | 15,797,436 | 0 | 100 | 100 | 3,182 | 75.3 |
| 9 | 781,633 | 1 | 99 | 99.9 | 3,381 | 80.0 |
| 8 | 385,359 | 4 | 96 | 99.8 | 3,479 | 82.3 |
| 7 | 391,535 | 4 | 96 | 99.7 | 3,571 | 84.5 |
| 6 | 342,124 | 12 | 88 | 99.5 | 3,638 | 86.1 |
| 5 | 293,419 | 6 | 94 | 99.4 | 3,717 | 88.0 |
| 4 | 258,017 | 20 | 80 | 99.1 | 3,773 | 89.3 |
| 3 | 342,086 | 54 | 46 | 98.2 | 3,812 | 90.2 |
| 2 | 348,753 | 51 | 49 | 97.2 | 3,851 | 91.1 |
| 1 | 395,202 | 74 | 26 | 95.8 | 3,874 | 91.7 |

According to the evaluation results in Table 26 for precision and recall, when the confidence level is between 4 and 10, the recall significantly decreases, while precision only shows a slight improvement. It appears that the accuracy of the PCS dataset on PubMed performs similarly to its performance on MAG. We agree with Matt's[74] suggestion that most scholars should focus on citations with a confidence level greater than or equal to 4. For those aiming to minimize erroneous data, citations with a confidence level of 10 can be used, as they exhibit both high precision and decent recall.

**Validation of paper-clinical trial citations**

We evaluated the linkage of clinical trial data through cross-validation with SciSciNet, as shown in Table 27. Due to differences in data time intervals, we validated the nct_id from the PKG dataset against SciSciNet.

Table 27 The accuracy of patent-paper linkage

| Paper-trial pairs | In PKG | Not in PKG |
|---|---|---|
| In SciSciNet | 428,627 | 4,376 |
| Not in SciSciNet | 76,154 | - |

There is a remarkable consistency between the two datasets, with 98.99% of the linkages in SciSciNet also present in PKG.

## Holistic validation of PKG 2.0

Holistic validation aims to assess the accuracy of interconnected linkages across papers, patents, and clinical trials. To validate the holistic accuracy of PKG 2.0, we conducted a manual verification process. For each project, we manually verified the linkages by accessing the original documents on PubMed, ClinicalTrials.gov, and USPTO's Patent Public Search platform (https://ppubs.uspto.gov/). The validation focused on citation relationships, and project-literature connections. For instance, among the randomly selected 100 patent-project linkages, upon checking on the USPTO's Patent Public Search platform, we found that 82 of them were indeed mentioned in the "GOVERNMENT INTEREST" section of the patent texts. By dividing the number of checked linkages (82) by the total number of linkages in the sample (100), we calculated the accuracy to be 82%. Regarding the performance of bioentity extraction, which also serves as a crucial linkage among papers, patents, and clinical trials, detailed information can be found in the "Results" section of the BERN2[59].

Table 28 The holistic validation of PKG 2.0

| Linkage Name | Accuracy |
|---|---|
| Link_Papers_ClinicalTrials | 100% |
| Link_Papers_Patents | 97% |
| Link_Papers_Projects | 99% |
| Link_ClinicalTrials_Projects | 98% |
| Link_Patents_Projects | 82% |

As shown in Table 28, the accuracy of these linkages ranged from 82% to 100%, with most exceeding 95%. However, patent-project linkages showed relatively lower precision (82%). This is likely because although the NIH Exporter might have included relevant data, the funding source information may not be mentioned in the patent texts. This validation confirms the robustness of PKG 2.0 in integrating multi-source biomedical literature and underscores its reliability for cross-domain scientometric analysis.

# Usage Notes

In the field of biomedicine, scholarly outputs are mainly presented in the form of clinical trials, patents, and academic papers. These different types of documentary data are complementary to each other. By comprehensively analyzing these literatures, it is possible to track the flow path of knowledge more accurately and deeply and then understand the characteristics and laws of knowledge flow. For example, the clinical trial results of vaccines and drugs may prompt patent applications and patent information may be introduced into papers to demonstrate their application prospects. This cross-literature knowledge flow not only helps scientists discover new research directions but also provides practical guidance for clinical doctors, ultimately promoting the overall progress of the medical field. This requires a multi-perspective, fine-grained connection to the three types of literature.

We attempt to explore the optimization of the multi-literature dataset and the analysis of knowledge flow through the following use cases using the PKG dataset: integrating features from patents and clinical trials into traditional author disambiguation methods; using COVID-19 vaccine research as a case study to transcend literature type limitations and more comprehensively explore knowledge flow.

## Enhancing Data Quality through Multi-source Fine-Grained Linkages

In academic datasets, author disambiguation has always been one of the key challenges. Although many algorithms and models have been proposed and implemented with relatively good results, these methods seem to have reached their performance limits. This study combines information from different sources and integrates and deeply correlates multiple datasets, which can more accurately capture the unique content features of authors, providing possibilities for further optimizing author disambiguation tasks.

Specifically, this study constructed pairwise experimental data from the Author-ity 2018 dataset based on authors with the same Initial, using whether they have the same links to patents or clinical trials as the standard. Name similarity, Publication year gap, Venue similarity, Affiliation similarity, and Content features commonly used in author disambiguation research were used as baseline model features (BF), and features with the same clinical trial links and the same patent links were subsequently added.

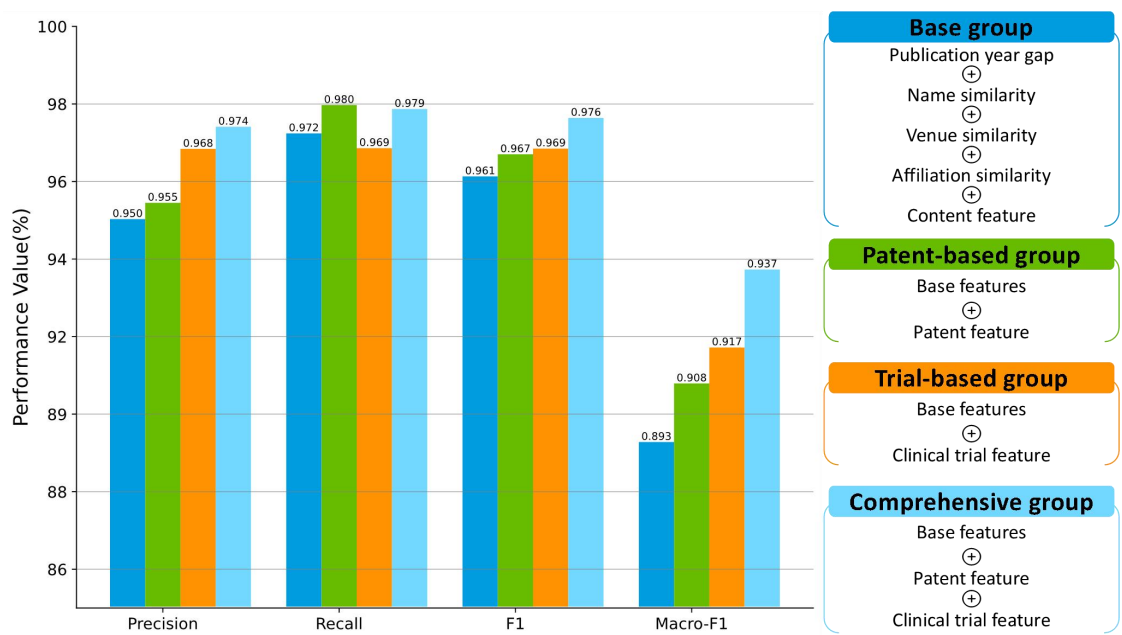

Fig. 4 Evaluation of author name disambiguation results for different feature groups. The Base group contains five commonly used features in the author name disambiguation task as a baseline; the Patent-based group and Trial-based group, respectively, incorporate features related to patents and clinical trial studies, which improve author disambiguation performance; after adding all features, the performance of the Comprehensive group reaches a relatively optimal level.

As shown in Figure 4, We report the accuracy (P), recall rate (R), F1, and Macro-F1 metrics based on the classification-based method. The reason for using the additional metric Macro-F1 is that the datasets in the author's disambiguation research generally have natural skewness. As shown in the table above, the disambiguation result has been greatly improved after adding the Clinical trial and Patent features. This example provides us with ideas for cross-iterating the overall optimization of the dataset using multiple types of data.

**Revealing knowledge flow through Multi-source literature**

The outbreak of the COVID-19 pandemic has not only posed unprecedented challenges to global public health but also sparked an unprecedented research boom in the academic community. The large number of patent applications, publication of papers, and conduct of clinical trials all represent scientific efforts to combat the pandemic. The rich literature resources have played a promoting role in research, and some scholars have conducted quantitative analysis and exploration of the literature. Despite the large number of literatures, most studies are based on a single type of literature resource, leading to an incomplete understanding of key areas such as vaccine and drug research. Analyzing multiple types of literature simultaneously, leveraging the PKG 2.0 dataset, can help to understand the development process of this research more accurately and reveal important synergies among different types of knowledge.

As a case study, we utilized the bioentity data and source institution data of the trials in PKG 2.0 to screen out the earliest ten COVID-19 clinical trials sponsored by BioNTech SE.

Then, by leveraging the linkages within PKG 2.0, we identified their earliest relevant papers and patents. As depicted in Figure 5, when these are plotted on the timeline of the same graph for analysis, the flow of knowledge across different types of literature becomes clearly visible.

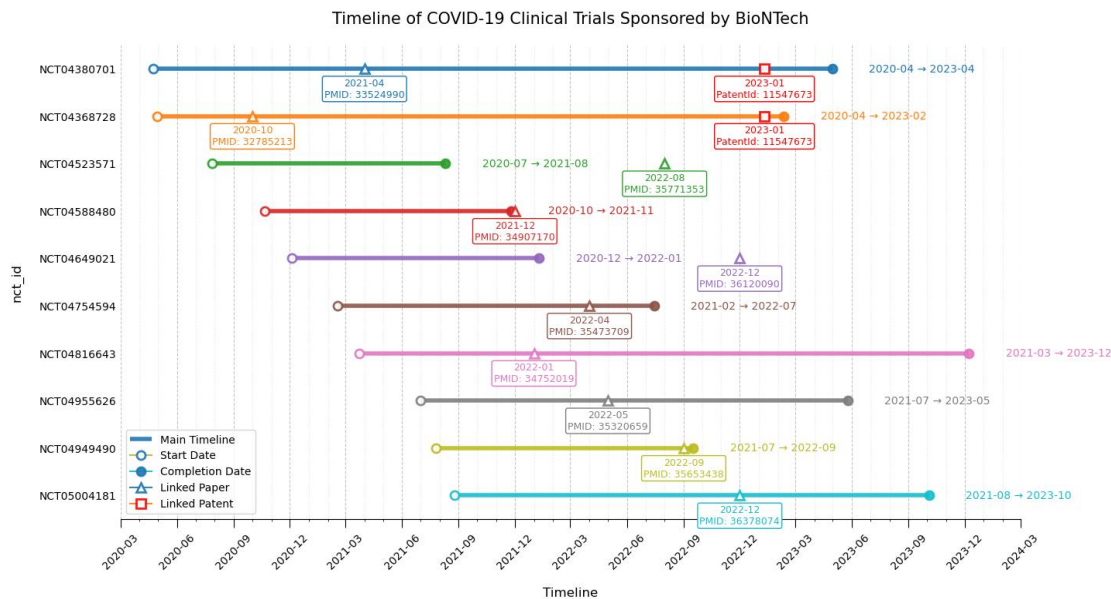

Fig. 5 Timeline of COVID-19 Clinical Trials by BioNTech. It showcases the earliest ten COVID-19 clinical trials sponsored by BioNTech. For each trial, a horizontal line represents its duration, stretching from the start date to the completion date. Along these lines, marker points are placed to denote the publication times of relevant papers and patents.

For a significant portion of these trials, the timeline indicates a relatively short interval between the start of the trial and the publication of related papers. This rapid dissemination of trial results through academic papers is crucial for sharing scientific findings within the research community. Moreover, the earliest trials in the dataset even led to patent applications. This example showcases a significant synergy that goes beyond simply pointing to individual papers or studies. It reveals how different types of knowledge from clinical trials, academic papers, and patents interact and build upon each other. By integrating and analyzing these three different datasets within PKG 2.0, we can understand not only the specific process of vaccine research but also the knowledge flow in a more comprehensive and accurate manner. For instance, we can observe how early clinical findings led to academic publications that then guided the development of patented vaccination techniques, highlighting the seamless flow of knowledge and innovation within the COVID-19 vaccine research domain.

Taking one of the COVID-19 vaccine research projects as an example, as shown in the Figure 6, its clinical trials for vaccine safety and immunogenicity began on April 29, 2020. The clinical trial results were reported in a paper published in Nature on August 12, 2020, followed by a patent application submitted to the USPTO on April 16, 2021 (granted on Jan 10, 2023). These three types of documents are interrelated: clinical trial reports cite papers, papers register clinical trial numbers, and there are citations among patents, papers, and clinical trials. Additionally, these documents often share common biomedical entities.

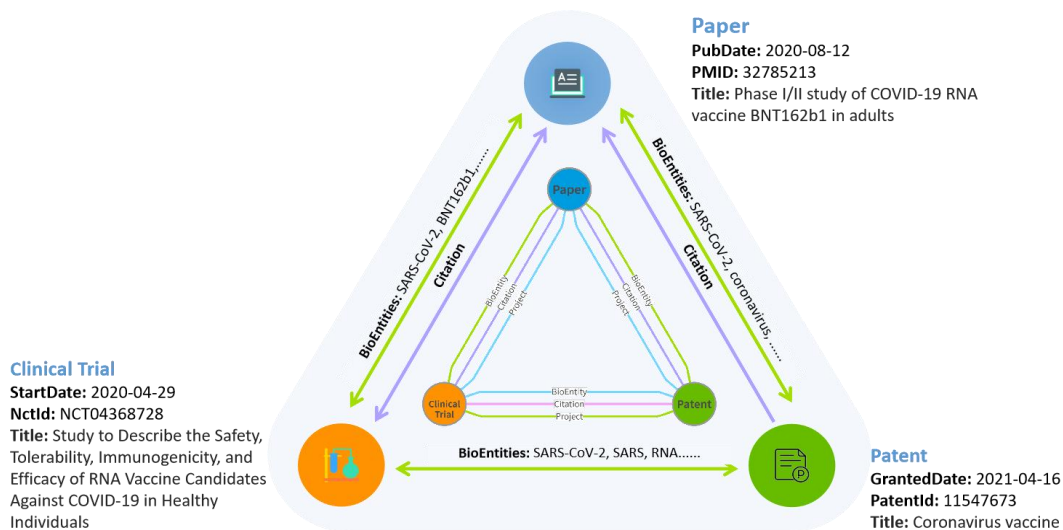

Fig. 6 The patent, paper, and clinical trial related to a certain COVID-19 vaccine. The internal triangle in the figure shows the linkages among the three types of literature in the PKG, and the outer one shows a specific instance of the COVID-19 vaccine. There are many related papers, clinical trials, and patents about the vaccine, and the figure only demonstrates a certain one of each type of literature. For example, a series of related papers from different time periods can be found through citations in clinical trials NCT04283461, including early reports focusing on vaccine safety and efficacy in various populations (PMID33053279/33301246/33524990), mid-term studies on vaccine effectiveness and adverse events (PMID35739094/35792746/36055877), and more recent reviews and exploratory studies (PMID38012751/35472295).

The clinical trial sponsored by BioNTech SE (ClinicalTrials.gov ID NCT04368728) was a comprehensive study evaluating multiple aspects of different RNA vaccine candidates against COVID-19. In this trial, various parameters like the safety, tolerability, immunogenicity, and efficacy of vaccine candidates were investigated. For instance, it explored different doses of vaccine candidates such as BNT162b1 and BNT162b2 in different age groups and also considered aspects like booster doses and potential protection against emerging variants.

This clinical trial was closely related to a paper published in Nature titled "Phase I/II study of COVID-19 RNA vaccine BNT162b1 in adults". The paper reported the available safety, tolerability, and immunogenicity data from an ongoing placebo-controlled, observer-blinded dose-escalation study, which was precisely the clinical trial with the identifier NCT04368728. The findings in the paper, such as the dose-dependent nature of local reactions and systemic events, and the increase in RBD-binding IgG concentrations and SARS-CoV-2 neutralizing titres after vaccination, provided crucial insights based on the clinical trial data. This synergy between the clinical trial and the paper shows how real-world data from the trial was analyzed and presented in an academic paper, contributing to the scientific understanding of the vaccine's characteristics.

Subsequently, a patent titled "Coronavirus vaccine" filed by BioNTech SE. This patent related to methods and agents for vaccination against coronavirus infection and inducing effective coronavirus antigen-specific immune responses. It built upon the knowledge gained from the clinical trial and the paper, translating the scientific findings about the vaccine's potential and its effects on immune responses into a protectable intellectual property related to vaccination methods.

Previous efforts in knowledge graphs, such as the Fine-grained single-type literature knowledge graph and Coarse-grained Integrated multi-type academic graph, have their limitations. For example, the Fine-grained single-type literature knowledge graph focuses on only one type of literature, if we were relying solely on a single-type literature knowledge graph, we would miss out on crucial connections. For instance, we wouldn't be able to see how the clinical trial results directly influenced the patent application or how the patent's proposed vaccination methods were rooted in the real-world data from the clinical trial and the theoretical basis provided by the academic paper. The Coarse-grained Integrated multi-type academic graph, although integrating multiple types of literature, lacks the fine-grained comprehensive links that PKG 2.0 offers. In the context of our COVID-19 vaccine example, it might show general associations between the clinical trial, paper, and patent but would fail to capture the detailed entity-level relationships. Through PKG 2.0, the linkages among these three types of documents become evidence. For example, in PKG 2.0, we can trace how specific bioentities, projects, and even the researchers involved were consistently represented and connected across the clinical trial, paper, and patent. The clinical trial results informed the content of the paper, which in turn influenced the technological aspects covered in the patent.

**Limitations and Future Directions for the PKG Dataset**

PKG 2.0 has several limitations. In terms of data access, currently, the knowledge graph is mainly offered in TSV and SQL formats. However, we are well aware that to more effectively meet the diverse needs of the research community, enhanced flexibility is indispensable. Bearing this in mind, we have plans to investigate alternative data access methods, such as the development of a user interface. This initiative is intended to streamline the data querying process, facilitating easier access for users, especially those without sophisticated technical capabilities, to explore the extensive information within the dataset. Furthermore, PKG2.0 dataset involves probabilistic and statistical models for tasks such as knowledge entity extraction, author disambiguation, and institutional disambiguation. While the current results represent the best achievable outcomes given the state-of-the-art methodologies, certain biases remain inevitable due to the inherent limitations of these models. We remain committed to the continuous improvement of these processes and will strive to further refine the dataset in future iterations.

The PKG dataset updating is a complex task that demands substantial computational resources and careful consideration of various factors. Firstly, it involves updates from multiple diverse data sources. Each update entails not only the downloading of data from these sources and authorization verification but also meticulous source data formatting and comprehensive quality checks. Additionally, significant code refactoring and data computation are required, which incur considerable computational costs. For example, the formatting process may involve converting data from different file formats and structures into a unified format suitable for integration into the PKG dataset, while the quality checks involve running multiple validation algorithms to ensure data integrity and accuracy. The code refactoring is essential to adapt to changes in the data sources and to optimize the processing pipeline for efficiency. As the dataset grows with each update and potential future expansions, the computational requirements will increase proportionally. Larger-scale applications that might utilize the PKG dataset, such as complex data analytics or machine learning models, will place even greater demands on computational resources. We need to consider how the current infrastructure and algorithms will scale to handle these increased loads.

In the future, we plan to update the dataset annually using the latest files from PubMed, PatentsView, ClinicalTrials.gov, and other related sources. To ensure ongoing data consistency and accuracy, we will implement a multi-step quality control process. Firstly, before each update, we will conduct a thorough review of the source data to identify and rectify any potential errors or inconsistencies. During the update process, we will use advanced data validation algorithms to cross-check and verify the new data against the existing dataset. After the update, we will perform post-update audits to confirm that the integrity of the dataset has been maintained. In addition to the annual updates based on PubMed, USPTO, and ClinicalTrials.gov update files and other reliable sources, we will establish a detailed monitoring mechanism. This mechanism will keep a close eye on significant changes in the biomedical field, such as emerging research trends, new data sources, and regulatory requirements. If there are any substantial developments that necessitate more frequent updates than the annual schedule, we will adjust the update frequency accordingly.

## Code Availability

The PubMed Knowledge Graph is available at https://pubmedkg.github.io, and the advanced neural biomedical named entity recognition and normalization tool is available at https://github.com/dmis-lab/BERN2.

## Acknowledgments

This work was supported by the National Natural Science Foundation of China [Grant Number: 72374233], Natural Science Foundation of Guangdong Province [Grant Number: 2024A1515011778]. We would like to express our sincere gratitude to the reviewers and editors for their meticulous evaluation and insightful suggestions. Their valuable feedback has significantly contributed to the improvement of this manuscript, helping us to refine our work and present a more comprehensive and accurate study.

## Author contributions

Jian Xu and Ying Ding proposed the idea and supervised the project.
Chao Yu and Jian Xu implemented the construction of the dataset and wrote and revised this manuscript.
Jiawei Xu and Yi Bu carried out the method and dataset entity extraction.
Vetle I. Torvik provided the Author-ity and MapAffil dataset, and guided the disambiguation process.
Jaewoo Kang, Mujeen Sung, and Min Song carried out the bio-entity extraction and validity.

# Competing interests

The authors declare that they have no competing interests with respect to this paper.